\documentclass{article}

\usepackage{PRIMEarxiv}

\usepackage{amsmath}
\usepackage{makecell}
\usepackage{bm}

\usepackage[utf8]{inputenc} % allow utf-8 input
\usepackage[T1]{fontenc}    % use 8-bit T1 fonts
\usepackage{hyperref}       % hyperlinks
\usepackage{url}            % simple URL typesetting
\usepackage{booktabs}       % professional-quality tables
\usepackage{amsfonts}       % blackboard math symbols
\usepackage{nicefrac}       % compact symbols for 1/2, etc.
\usepackage{microtype}      % microtypography
\usepackage{lipsum}
\usepackage{fancyhdr}       % header
\usepackage{graphicx}       % graphics
\graphicspath{{media/}}     % organize your images and other figures under media/ folder

\title{Analysis of Proximity Informed User
Behavior in a Global Online Social Network
\thanks{The authors are listed in alphabetical order by last names, and all contributed equally to this work, sharing first authorship.}
}

\author{
  Nils Breitmar \\
  Goethe University Frankfurt \\
  Frankfurt, Germany\\
  \texttt{breitmar@wiwi.uni-frankfurt.de} \\
   \And
  Matthew C. Harding \\
  University of California Irvine \\
  Irvine, CA\\
  \texttt{matthew.harding@uci.edu} \\
  \AND
  Hanqiao Zhang \\
  University of California Irvine \\
  Irvine, CA \\
  \texttt{hanqiaoz@uci.edu} \\
}

\begin{document}
\maketitle

\begin{abstract}
Despite the earlier claim of "Death of Distance", recent studies revealed that geographical proximity still greatly influences link formation in online social networks. However, it is unclear how physical distances are intertwined with users' online behaviors in a virtual world. We study the role of spatial dependence on a global online social network with a dyadic Logit model. Results show country-specific patterns for distance effect on probabilities to build connections. Effects are stronger when the possibility for two people to meet in person exists. Relative to weak ties, dependence on proximity is looser for strong social ties.
\end{abstract}

% keywords can be removed
% \keywords{Social Network \and Dyadic Logit}

\section{Introduction}

Online social networks are a microcosm of our increasingly globalized world, connecting individuals across vast distances and cultural divides. The structure of these networks, particularly the mechanisms that drive connection and people making new friends, are areas of ongoing research interest. As we delve into the digital age, the importance of geographical proximity, a key player in traditional, offline networks, warrants close examination in the online context.

Many proposed friend recommendation algorithms for online social networks consider physical proximity as an important factor for link formation, e.g. see \cite{xie2010potential, chin2013should, chin2012linking,wang2011interplay}. This is based on the belief that geographical closeness can be a proxy for a higher likelihood of connection. However, other factors, such as users' ethnic group, native language, shared interests and experiences, may also dictate the preference to select whom to follow. Therefore, evaluating the significance of physical proximity in digital spaces could be very helpful in refining friend recommendation algorithms to enhance efficiency. Moreover, for global social networks, the investigation should be conducted separately with respect to users from different countries, because the importance of geographical proximity might differ due to differences in culture, regulatory environment, demographics, etc. Comparative studies have been covered in previous research, but they were typically confined to a comparison between two or three countries, relying largely on subjective self-reported data. This paper embarks on a more comprehensive exploration of the role that geographical proximity plays in the formation of digital connections, particularly in the wake of the 2020 pandemic and subsequent quarantine regulations, which catalyzed an unprecedented reliance on virtual modes of interaction.

For the rest of this paper, section two reviews related literature and generates hypothesis, section three introduces the social network dataset and the model, section four presents empirical results, and the last section concludes.

The study of how spatial propinquity affects the formation of social ties dates back to the 1950s, long before the advent of the Internet, in the context of face-to-face contact and phone calls. Evidence shows that physical space played an essential role and is inversely proportional to interaction frequency \cite{snow1981social,blake1956housing,latane1995distance}. Information technology reshapes the structure of the social network by greatly reducing the cost of making new friends and maintaining relationships, leading to the statement of "Death of Distance" in \cite{cairncross2001death}. It is conceivable that in the modern era, geographical location becomes less of a constraint for building connections, but there is still the desire for people to meet each other and have a cup of coffee physically. For example, \cite{goldenberg2009distance} argues that the ease of communication further strengthens local social ties, and the communication volume is inversely proportional to people's geographic distance. \cite{boase2006article} finds that people who send more emails to each other also have more frequent face-to-face and phone contact.
There are many other pieces of literature that demonstrate the tendency to form short-distance links or geographically closed friendship clusters \cite{scellato2010distance, liben2005geographic, hipp2009simultaneous, backstrom2010find}. It leads to our first hypothesis with respect to the influences of proximity on online link formation:

\textit{H1. The possibility of a user following another user declines along with the growth of their geographical distance.}

Although online social networks exhibit distance dependence like offline, the spatial dependence may decay in different patterns or at a slower rate. This is because the cost of building online connections (clicking the "follow" button on someone's profile) is much cheaper than breaking the ice and getting acquainted with other people face-to-face. As a result, the network is usually composed of both strong and weak ties, leading to a lower strength of connections. Besides keeping in touch, people have a variety of reasons to follow other people's lives: resonate with others' thoughts, be attracted by posted pictures, or simply follow the crowd. This could be particularly true for weak ties in online directed networks, like Instagram or Twitter. Users follow others to learn new knowledge, seek values to help improve their lives or realize the lifestyles of their dreams. Thus it is possible that in the digital space, the appeal of popularity transcends geographical boundaries, leading users to form connections based on interests, influence, and cultural factors. The simplest hypothesis to test is to see whether users are more likely to follow others from popular places rather than those who are geographically close, especially when they could hardly meet in person:

\textit{H2. People are more likely to follow others from popular countries.}

This could be due to several reasons. Popular countries often have a significant cultural influence that extends beyond geographical boundaries. For example, users from Italy or France may be seen as more authoritative in fields like fashion due to the perceived status of these places. Users may follow individuals from these places to keep up with trends and ideas, and have exposure to a wide array of opinions, lifestyles, and experiences, thereby enriching their own social media feeds with diverse content. Popular places may also cultivate more Internet celebrities that attract followers.

In addition, online social networks, especially global ones, comprise far more entities than offline communities from all countries and have multitudinous cultural milieus and diverse lifestyles. Research shows people from different countries have distinct motivations for using online social networks, e.g., \cite{kim2011cultural} found that Korean students attached more importance to gaining social support from their friends, and students from the United States put a larger weight on entertainment. Different motivations may yield divergent social behaviors. \cite{cardon2009online} surveyed university students from 11 countries, showing that users from collectivist nations built more connections with whom they had never met. In contrast, users from individualist nations nurture more relationships offline. In other media, \cite{chen2002global} investigated people who connect with their relatives face-to-face, by telephone, and by email, reported heterogeneity within and beyond 50 kilometers in North America, other developed countries, and developing countries. These form the basis of our third hypothesis:

\textit{H3. There is heterogeneity in how users' online behaviors are affected by proximity in different regions. }

Another angle that supports country heterogeneity is provided by papers that study the variation in network structures following geographical variability and unevenly distributed populations. Inhabited areas are usually rich in resources, better in climate, or have other desirable properties. Barren lands and oceans, where there are few or no users at all, could certainly affect the estimated effect of region-specific geographical proximity, e.g., see \cite{butts2012geographical}.

\section{Dyadic Logit Model with Social Network Data}

The social network data is collected through three processes by the mobile app: user registration, user profile, and user activity. As the first point of contact with the platform, users are asked to provide basic information about themselves during registration, such as name, email address, gender, and date of birth. In the second step, users are encouraged to complete their profiles by adding more personal information, such as their heights and weights, ethnicity, relationship status, and profile pictures. When users start to interact with the app, their online activities are logged as user behavioral data, such as the followers and followees, their liked and commented posts and stories, chat messages that are sent, etc.

We focus on a snapshot of the online social network, on December 16th, 2020, that consists of 11,992 active users, stratifed by country of origin, who have clicked the "follow" or "unfollow" button at least 4 times a day in the sample. Inactive users are not considered, because their actions are not representative of user behaviors in the network, and may obscure the relationship between geographical distance and online interactions.

Similar to other large-scale social networks in reality, the directed network is sparse with 0.59\% of the total number of edges possible actually present. The in-degree distribution follows a power law of the form: $P(d) \propto cd^{-\lambda}$, where $d$ is the vertex's in-degree, if plotted to logarithmic scale, see \cite{faloutsos1999power,albert1999diameter}. The power law produces a highly skewed histogram, where very few users are celebrities with plenty of followers, and most of the users are sparsely linked. The median user has 40 followers, and 0.12\% of the users own 1,000 followers.

\begin{figure} \centering
  \caption{Vertex In-degree Distributions}
  \begin{minipage}[t]{0.45\textwidth}
    \centering
    \includegraphics[width=7cm]{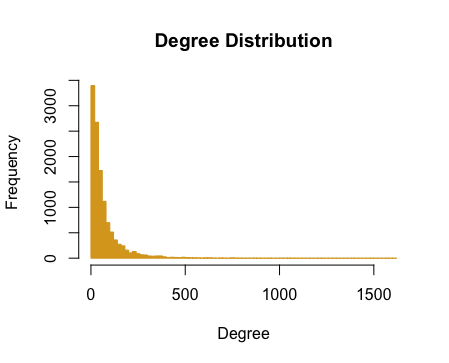}
  \end{minipage}
  \begin{minipage}[t]{0.45\textwidth} \centering
    \includegraphics[width=7cm]{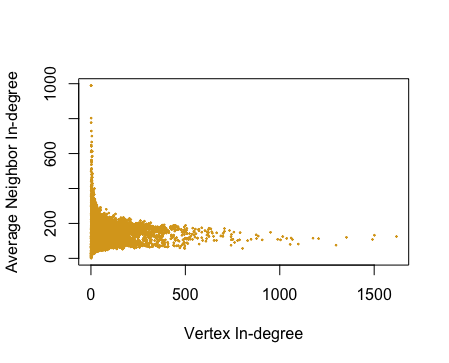}
  \end{minipage}
\end{figure}

Beyond the degree distribution, how users with different numbers of fans are linked could be seen by plotting the vertex in-degree versus the average in-degree of its neighbor. Users with few or no followers are more exploratory in their connections, as they are still trying to build their social network. They may connect with celebrity users due to their popularity or influence, while also following low-degree users who are closer to their own levels of influence. When users have more followers, their neighbors' average in-degree uncertainty decreases. This may imply that vertices with higher degrees are more selective about who they follow, and they tend to connect with other high-degree nodes, leading to a more predictable degree distribution among their neighbors. This could be due to a number of reasons, such as their desire to maintain a certain image, the limited time they have to interact with their followers, or their strategic goal to maximize their influence or reach.

\begin{figure} \centering
  \caption{Social Network Visualized with Users' Geographical Locations}
  \includegraphics[width=\textwidth]{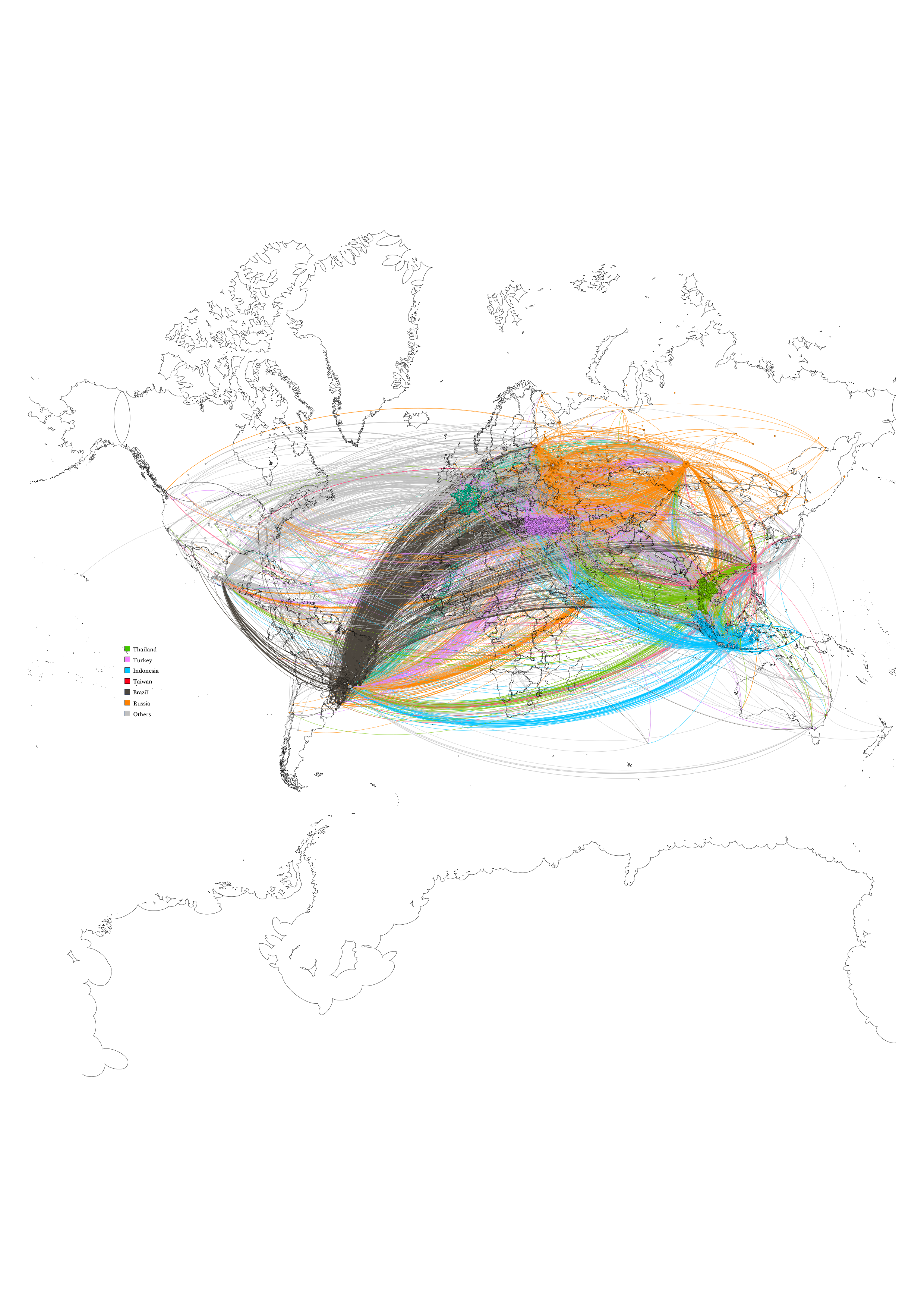}
\end{figure}

We visualize the network by setting the vertices' size to correspond to the user's followers, and the vertex color to the user’s country of origin. There is an edge between any two users if one user follows the other, and the colors of the edges accord with that of the followers. The user base covers all five continents and a few islands but is mainly located in Asia, Europe, and South America. For countries, 28.67\% users come from Thailand, 27.65\% users are from Turkey, 10.66\% users are from Indonesia, 9\% users come from China (Taiwan), 7.48\% are from Brazil, and 5.57\% users come from Russia. The visualization shows that the network is truly global: people follow others from different countries or continents.

Due to the measuring scale, it is hard to see how the density of connections changes along with distance, so We compare the histogram of the distance between any two users, and the histogram of any two mutually-connected users. Distance between any two users ranges from 0 to 20,000 kilometers. The distribution has three peaks: less than 100 kilometers, near 2,500 kilometers, and around 7,000 kilometers. These may correspond to three typical scenarios: two users from the same area, two cities of the same country, and two cities of different countries. For these three groups of users, mutual connections are most easily formed for those in the first group, when two users are from the same area.

\begin{figure} \centering
  \caption{Histograms of the Distance between All Users and Mutually Connected Users}
  \begin{minipage}[t]{0.45\textwidth}
    \centering
    \includegraphics[width=7cm]{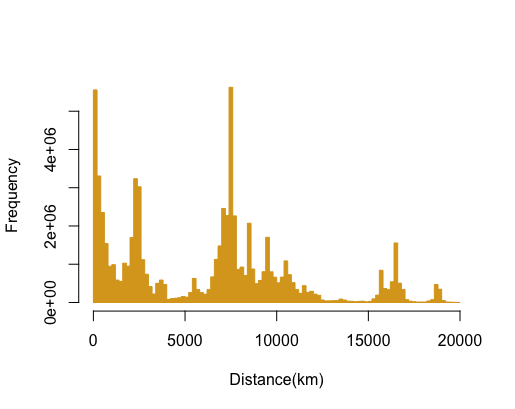}
  \end{minipage}
  \begin{minipage}[t]{0.45\textwidth} \centering
    \includegraphics[width=7cm]{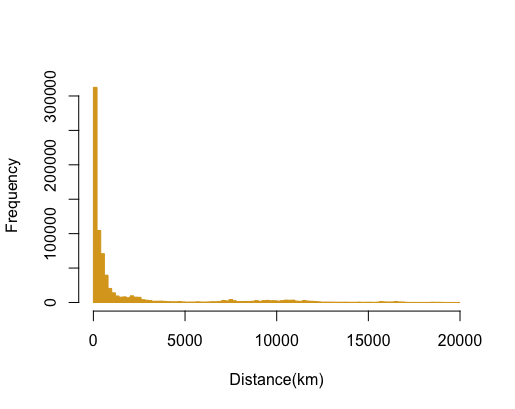}
  \end{minipage}
\end{figure}

To explore the user preference for following others among different countries, we produce the heatmap, for the 10 largest countries, by listing followers' nationalities on the y-axis, and followees' nationalities on the x-axis. Users from Thailand, Turkey, Indonesia, and China (Taiwan) mainly build connections internally. Brazilian and Russian users also prefer to follow others from the same country but are open to following users from other places. More diversities are seen among people from Ukraine, Mexico, and the United States. The heterogeneity could be jointly induced by a variety of cultural, linguistic, and sociopolitical factors. For example, users are more likely to follow others who speak the same language and share the same cultural background. Higher diversity in countries like the U.S. may be contributed by higher internet penetration, the larger size of the online community, the presence of more diaspora communities, etc.

\begin{figure} \centering
  \caption{Heatmap of Users' Countries of Following}
  \includegraphics[width=0.45\textwidth]{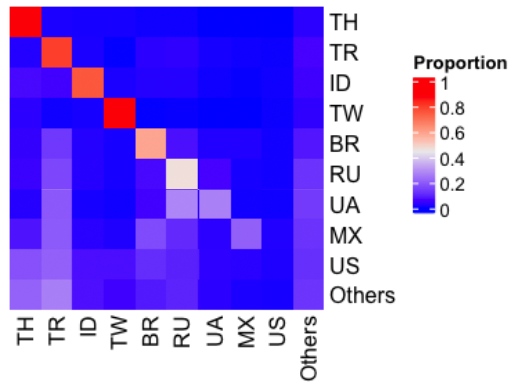}
\end{figure}

In \cite{tinbergen1962shaping}, dyadic regression is initially used to model the logarithm of exports from country $i$ to country $j$ with logged Gross National Product (GNP) of both countries, logged distance of the two countries and other variables. After that, the setting is largely applied to empirical works in international trade, e.g., \cite{anderson2011gravity, rose2004we}. Other fields in social sciences, e.g., \cite{portes2005determinants} explains bilateral financial assets transactions among 14 countries from 1989 to 1996, \cite{atalay2011network} characterized the buyer–supplier network in the U.S., \cite{owsiak2021peaceful} investigates how peaceful country dyads are formed from a territorial perspective, just to name a few. To estimate the influence of geographical proximity on users' probabilities to connect, we implement a series of dyadic Logit regressions to model the data. Following the notation from \cite{de2020econometric}, a simple dyadic Logit model, for either a directed or undirected network, could be specified as:
\begin{equation}
  \nonumber
  W_{ij}=1\left\{ D_{ij} \beta_D + X_{ij}^T \beta_X + \epsilon_{ij} \geq 0 \right\}
\end{equation}
where $W_{ij}$ is the connectedness of user $i$ and $j$, $D_{ij}$ are pair-wise covariates of interests, such as geographical distance between the two users, $X_{ij}$ includes control variables which could be user-specific or dyad-specific variables, and $\epsilon_{ij}$ is a logistic random variable. We include a rich set of vertex features as well as a series of constructed dyadic features that reflect homophily in $X_{ij}$. Summary statistics of the vertex attributes are shown in the following table:

\begin{table}[!htbp] \centering
  \caption{Summary Statistics of Vertex Attributes}
\begin{tabular}{lcccc}
\\[-1.8ex]\hline  \hline \\[-1.8ex]
Statistic & \multicolumn{1}{c}{Mean} & \multicolumn{1}{c}{St. Dev.} & \multicolumn{1}{c}{Min} & \multicolumn{1}{c}{Max} \\
\hline \\[-1.8ex]
App Usage in Days & 662.70 & 285.990 & 4 & 910 \\
App Version & 4.72 & 0.87 & 3 & 7 \\
Platform IOS & 0.25 & 0.43 & 0 & 1 \\
Age & 30.253 & 9.95 & 18.00 & 99.00 \\
Height(cm) & 181.69 & 153.14 & 91.00 & 200.00 \\
Weight(kg) & 70.79 & 14.23 & 27.00 & 293.00 \\
Uploaded Photo & 0.82 & 0.39 & 0.00 & 1.00 \\
Account Visible & 0.97 & 0.17 & 0.00 & 1.00 \\
Has Email & 0.70 & 0.46 & 0.00 & 1.00 \\
Photos Rejected by Platform & 6.00 & 26.47 & 0 & 1,477 \\
Feed Posts Made in V4 & 0.63 & 53.77 & 0 & 5,872 \\
Feed Posts Made in V5 & 7.55 & 49.70 & 0 & 1,996 \\
Feed Posts Made in V6 & 14.11 & 70.64 & 0 & 2,744 \\
Feed Posts Liked/Commented in V4 & 44.94 & 1,101.46 & 0 & 84,481 \\
Feed Posts Liked/Commented in V5 & 789.14 & 4,130.66 & 0 & 116,539 \\
Feed Posts Liked/Commented in V6 & 1,079.40 & 4,214.73 & 0 & 205,533 \\
Stories Read from Feed in V4 & 0.05 & 0.83 & 0 & 56 \\
Stories Read from Feed in V5 & 47.40 & 228.67 & 0 & 10,625 \\
Stories read from Feed in V6 & 64.41 & 240.61 & 0 & 14,292 \\
Chat Messages Sent in V4 & 365.88 & 2,498.43 & 0 & 87,302 \\
Chat Messages Sent in V5 & 5,047.41 & 11,383.11 & 0 & 175,948 \\
Chat Messages Sent in V6 & 6,181.82 & 8,414.75 & 0 & 116,155 \\
Total Followees in V4 & 50.51 & 525.83 & 0 & 22,546 \\
Total Followees in V5 & 842.16 & 3,080.15 & 0 & 167,086 \\
Total Followees in V6 & 1,166.10 & 1,656.27 & 0 & 49,161 \\
Total Followers & 70.21 & 97.15 & 0 & 1,618 \\
Total Followees & 70.21 & 115.53 & 0 & 4,243 \\
\hline \hline \\[-1.8ex]
\end{tabular}
\end{table}

The sample covers both new users who register an account as late as 4 days ago and old users who have had the account for more than 2 years, including all age groups. Among those who report their races, 47.86\% are Asian, 24.40\% are White, 7.47\% are Mixed, and 20.27\% are from other ethnic groups. Self-reported first languages largely coincide with users' countries of origin. 26.76\% users speak Turkish, 25.78\% speak Thai, 9.77\% speak Bahasa Indonesia, 8.83\% speak Mandarin, and 8.04\% speak English. Users also exhibit different social behaviors. Some lurkers remain inactive on the platform, while others actively send chat messages or interact with a few thousand posts in their feeds. Users' followers range from 0 to 1,618, and the median user has 40 fans.

To investigate the second hypothesis, we define the popularity index for the country as the number of followers from other countries normalized by the number of users from that country, and include it as an additional regressor:

\begin{equation}
  \nonumber
  Popularity = \frac{Number\ of\ Followers\ Outside\ the\ Country}{Country\ User\ Base}
\end{equation}

The country popularity index stands for how many times the number of followers outside the country is to the size of the user base or the average number of foreign followers each user from that country has. Country popularity ranges from 0 to over 80, and the smaller the size of the user base, the more popular the country is. We restrict the user base to be larger than 50 and define the 7 countries with the highest popularity index as popular: Germany, France, United States, Mexico, Ukraine, Iran, and Russia.

\begin{figure} \centering
  \includegraphics[width=0.5\textwidth]{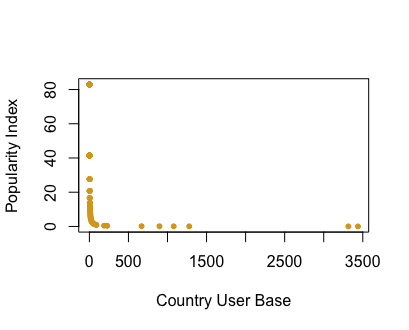}
    \caption{Country Popularity}
\end{figure}

We apply the model for both the weakly-tied directed network, in which $W_{ij}=1$ if user $i$ follows $j$, and the strongly-tied friendship network, in which $W_{ij}=1$ only if user $i$ and $j$ have followed each other. Apart from the models for all users, to investigate country heterogeneity, we run country-specific models for $6$ countries with the largest user base, by restricting the nationality of the follower of each dyad.

Although the dyadic Logit model assumes independence for all dyads and overlooks the strategic part of building connections, it could still replicate stylized attributes of social networks, see \cite{jochmans2018semiparametric}. The challenge that prevents more advanced settings from being adopted is mainly computational. $11,992$ users generate a sample size of more than $143$ million dyads, each with more than $100$ control variables. The data size makes it impossible to fit into computer memory at one time, and to our best knowledge, there is no off-the-shelf package that runs Logit models at this scale. Thus we used a single-layer neural network, and train it with one smaller batch of the dataset at a time to estimate the parameters.

It is worth mentioning that nationwide travel restrictions and region or city-level stay-at-home orders were enforced for different countries in December 2020. This would rise the probability of people registering and using online social networks, affect the compositions of the user population, or potentially change their social behaviors compared to pre-pandemic times. In addition, 7.14\% of users in the network live in Russia and Ukraine. Although the Russo-Ukrainian War has not been escalated until February 2022, the conflict between the two countries, which may include troop incidents, cyber-attacks, and political tensions, could date back to 2014. Online behaviors of users in the area may be affected by following or unfollowing users from other countries.

\section{Empirical Results}

We first present the Logit coefficients for the friendship network, on which only mutual connection between any two users is considered an edge. The model is run for users from the world at first, then for users from the 6 largest countries respectively by restricting followers' countries of origin: Thailand, Turkey, Indonesia, China (Taiwan), Brazil, and Russia.

\begin{figure} \centering
  \includegraphics[width=0.9\textwidth]{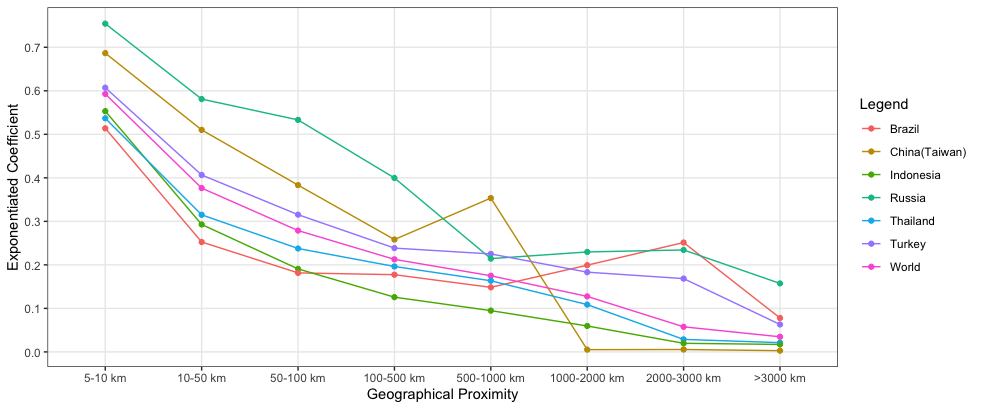}
    \caption{Exponentiated Coefficients of Friendship Network by Country}
\end{figure}

For geographical proximity, the reference level for all models is set to those pairs of users whose distance is within 5 kilometers. Proximity is still of great importance in online social networks, as shown by the significant coefficients, and users are most likely to establish friendships with their neighbors. When the distance between two users becomes further but is still within the practical range to meet offline on a regular basis (100 kilometers), the probability of forming mutual connections declines. Different user behavior patterns could be observed for different countries when the distance between two users goes beyond 100 kilometers. For China (Taiwan), the probability drops until 500 km, then rises a little when the distance is between 500 and 1,000 kilometers. However, this upswing is not necessarily due to changes in user behaviors but rather the lack of users in the sample, given that China (Taiwan) is an island surrounded by the western Pacific Ocean. The connections are mainly built with users from Hong Kong. Taiwanese hardly establish strong ties with users more than 1,000 kilometers away. Another pattern is found for users from the largest country in the world, Russia: users are the most tolerant to longer distances, compared to other countries, within 100 kilometers, but the probability of friendship formation drops significantly if distances get further and catch up with the other countries near 1,000 kilometers. When distance surpasses 1,000 kilometers, the probability of mutually connecting with others grows slowly until the 3,000-kilometer threshold is reached. Besides those who are from another far-off Russian city, plenty of links are formed with users from Turkey. Correspondingly, geographical proximity plays a weaker role for Turkish users who are not from the same place. The probability of making friends remains the same for pairs of users within the range of 100 and 1,000 kilometers and of 1,000 and 3,000 kilometers. The only country in South America, Brazil, shows particular characteristics of how users use the social network app. Within shorter distances, the probability of establishing mutual connections declines sharply in the wake of growing distance, with the largest decreasing amplitude among all countries. When the two users have no chance to meet in person, the probability of following each other does not descend further and even rises again over a long distance. The other two Asian countries, Indonesia and Thailand, behave similarly: friendship gradually and steadily becomes more difficult to build between two people and the rising distance between them.

\begin{table}[!htbp] \centering
  \caption{Logit Coefficient of Countries' Friendship Network}
\begin{tabular}{lccccccc}
\\[-1.8ex]\hline \hline
\\[-1.8ex] & World & Thailand & Turkey & Taiwan & Indonesia & Brazil & Russia   \\
\hline \\[-1.8ex]
 Same Platform & 0.16$^{***}$ & 0.11$^{***}$ & 0.28$^{***}$ & 0.01 & 0.21$^{***}$ & 0.28$^{***}$ & 0.30$^{***}$ \\
    & (0.00) & (0.00) & (0.01) & (0.01) & (0.01) & (0.01) & (0.01) \\
    Same Continent & -0.73$^{***}$ & -0.70$^{***}$ & -0.99$^{***}$ & -0.68$^{***}$ & -0.70$^{***}$ & 1.83$^{***}$ & 1.15$^{***}$ \\
    & (0.01) & (0.01) & (0.01) & (0.02) & (0.01) & (0.02) & (0.01) \\
    Same Country & 1.04$^{***}$ & 1.44$^{***}$ & 1.41$^{***}$ & 0.07$^{***}$ & 1.18$^{***}$ & 0.03 & 0.20$^{***}$ \\
    & (0.01) & (0.01) & (0.01) & (0.02) & (0.02) & (0.05) & (0.02) \\
    Same Region & 0.34$^{***}$ & 0.91$^{***}$ & 0.11$^{**}$ & -0.03$^{**}$ & 0.06$^{***}$ & -0.08$^{***}$ & 0.57$^{***}$ \\
    & (0.01) & (0.02) & (0.05) & (0.01) & (0.01) & (0.02) & (0.07) \\
    Same City & -0.35$^{***}$ & -0.97$^{***}$ & -0.08$^{*}$ & 0.47$^{***}$ & 0.13$^{***}$ & 0.15$^{***}$ & -0.71$^{***}$ \\
    & (0.01) & (0.02) & (0.05) & (0.05) & (0.02) & (0.02) & (0.08) \\
    Age within 5 Years & 0.15$^{***}$ & 0.15$^{***}$ & 0.13$^{***}$ & 0.11$^{***}$ & 0.14$^{***}$ & 0.08$^{***}$ & 0.10$^{***}$ \\
    & (0.00) & (0.00) & (0.01) & (0.01) & (0.01) & (0.01) & (0.01) \\
    Same Ethnicity & 0.35$^{***}$ & 0.16$^{***}$ & 0.06$^{***}$ & 0.20$^{***}$ & 0.24$^{***}$ & 0.38$^{***}$ & 0.17$^{***}$ \\
    & (0.00) & (0.00) & (0.01) & (0.01) & (0.01) & (0.01) & (0.01) \\
    Height within 5cm & 0.25$^{***}$ & 0.28$^{***}$ & 0.08$^{***}$ & 0.21$^{***}$ & 0.10$^{***}$ & 0.13$^{***}$ & 0.08$^{***}$ \\
    & (0.00) & (0.00) & (0.01) & (0.01) & (0.02) & (0.01) & (0.01) \\
    Weight within 5kg & 0.10$^{***}$ & 0.18$^{***}$ & 0.01 & 0.07$^{**}$ & 0.04 & 0.10$^{**}$ & 0.12$^{***}$ \\
    & (0.01) & (0.02) & (0.02) & (0.03) & (0.06) & (0.05) & (0.03) \\
    Same Language & 0.31$^{***}$ & 0.24$^{***}$ & 0.31$^{***}$ & 0.10$^{***}$ & 0.44$^{***}$ & 0.14$^{***}$ & 1.05$^{***}$ \\
    & (0.00) & (0.00) & (0.01) & (0.02) & (0.01) & (0.04) & (0.02) \\
\hline \hline \\[-1.8ex]
\textit{Note:}  & \multicolumn{7}{r}{$^{*}$p$<$0.1; $^{**}$p$<$0.05; $^{***}$p$<$0.01} \\
\end{tabular}
\end{table}

To transform the scale from log odds to probability, we compute the marginal effects of geographical proximity at mean, for a typical dyad of each country. At the world level, compared with two users who live within 5 kilometers, the probability of the dyad connecting decreases 0.028 when the distance is between 5 and 10 kilometers, and 0.043 when the distance is between 10 and 50 kilometers. Probability reduces along with the increase in distance, and it decays slower for further distances. The table of the marginal effects is shown in \ref{marginal_effect_friendship}.

For covariate effects, users who use the same platform (IOS/Android) have a higher likelihood to connect. This correlation is found to be significant globally and within countries, with the coefficients ranging from $0.11$ to $0.30$, notably high in Turkey, Brazil, and Russia. This may be due to users who share the same platform having similar online social behaviors or belonging to similar demographic groups.

Residing on the same continent surprisingly reduces the odds of connection among users, except in Brazil and Russia. Looking more closely at national boundaries, being from the same country significantly increases the odds of connection, with particularly strong effects noted in Thailand, Turkey, and Indonesia. Region and city-level analysis produce mixed outcomes. While being in the same region generally leads to a higher probability of connection, the effects of being in the same city vary by country. These findings imply a nuanced and potentially complex relationship between geographical proximity and online social networking. The contrasting patterns among various places may reflect diverse social dynamics, such as varying urbanization rates or digital penetration levels.

Apart from geographical homophily factors, physical and cultural attributes also play a role. For instance, age differences of fewer than 5 years, sharing the same ethnicity, and being within 5 centimeters of height all positively influenced the likelihood of connection. The effect was most pronounced when the users speak the same language, particularly in Russia, indicating the vital role of the language barrier in online social interactions.

\begin{figure} \centering
  \includegraphics[width=0.9\textwidth]{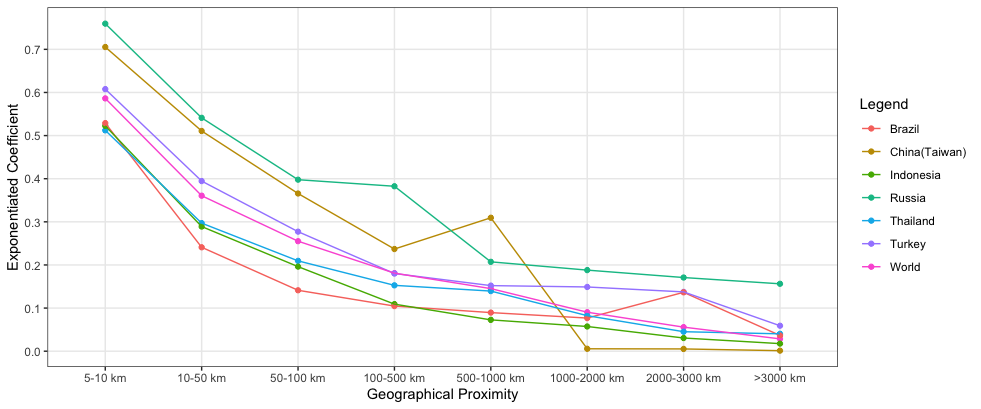}
    \caption{Exponentiated Coefficients of Directed Network by Country}
\end{figure}

When it comes to weaker social ties in a directed follow-unfollow network, country heterogeneity is observed as well. For the three Asian countries, China (Taiwan), Indonesia, and Thailand, probabilities of following another user in response to longer distances change in almost identical fashions with that in a friendship network, albeit with slightly larger distance decays. In Europe, the odds of Russian users following others now also change monotonically with proximity. $1,000$ kilometers is still a threshold near which the probability sharply declines. In South America, users from Brazil still treasure the opportunity to meet up very much whenever possible within a short range. However, when it is infeasible with distances further than $100$ kilometers, Brazilians are also interested in forming weaker connections by following users from other places.

Surprisingly, coming from popular countries seems to have no impact or negative impacts on the likelihood of being followed. Users tend to follow others of similar ages, and follower being older tends to increase the likelihood of a follow. Being in the same ethnicity generate higher probability to connect in Thailand, China (Taiwan) and Brazil, but not in other countries. People like to connect with others of similar heights, but the effect is not significant for building connections with users of similar weights. Preference for the height of the followee differs across countries. Taller users, compared to the followers, are more likely to be followed in China (Taiwan), Indonesia, and Brazil, but are less likely to be followed in Thailand and Turkey. User activities in the app, particularly in terms of posts made and stories read, vary widely in their impacts on the connecting likelihood.
The coefficients are shown in appendix \ref{logit_directed_coef_geo}, \ref{logit_directed_coef_homophily}, \ref{logit_directed_coef_cov_1}, \ref{logit_directed_coef_cov_2}.

In both the friendship and directed online social networks, users appreciate geographical proximity, and are most likely to connect with others who are within 5 kilometers:

\textit{Finding 1. Geographical proximity plays an important role in online social networks, despite the occurrence of country-specific patterns. }

Although connections are hard to establish with distance for all countries in general, the decay of odds to form social ties is slower for the friendship network, especially when the distance is beyond 500 kilometers, particularly for Russia and Brazil:

\textit{Finding 2. Strong social ties (mutual connections) are less sensitive to the change of physical distance than weak ties (unidirectional connections). }

Relative to the unidirectional relationship of a follower and a followee, users who follow each other are tended to send more chat messages, read each other's posts, share common values, and thus yield stronger bounds that help bridge the gap induced by geographical locations.

It is reported that people desire to catch up face-to-face or hang out together from time to time so that the majority of edges cluster within closed areas, e.g., see \cite{ugander2011anatomy, ellison2006spatially}. Nevertheless, it is unclear if the intensity of spatial dependence would remain the same at longer distances. In this data, the effects of proximity are lessened for pure net friends who are not able to meet offline, as depicted by the flattening slopes of all curves:

\textit{Finding 3. Spatial dependence is stronger when two users may meet in person and becomes looser when distances are longer. }

As pointed out in \cite{graham2020dyadic}, the dyadic Logit model assumes independence across dyads even when they share one or two users in common. It does not usually hold in reality, and uncontrolled interdependence may lead to standard errors and hinder the identification of coefficients. For example, apart from observed covariates, many unobserved user characteristics have not yet been controlled, such as personality, cultural background, etc. It is possible that these factors may also affect link formation, induces correlations among dyads, and thus obscure the identification of coefficients for geographical proximity, e.g., outgoing people may send friend requests to others indifferently regardless of their locations. To allow for correlated dyads and control for unobserved user heterogeneity, \cite{charbonneau2017multiple} and \cite{graham2017econometric} incorporate fixed effects into the model for directed and undirected networks, respectively. Borrowing the idea from panel data literature, the estimators are able to difference-out the unobserved individual fixed effects by focusing on tetrads that satisfy certain conditions.

To illustrate the method, assume user $i$ follows $j$ based on the following rules in a directed network:
\begin{equation}
  \nonumber
  W_{ij}=1\left\{ D_{ij} \beta_D + X_{ij}^T \beta_X + \alpha_i^{out} + \alpha_j^{in} + \epsilon_{ij} \geq 0 \right\}
\end{equation}
where $\alpha_i^{out}$ and $\alpha_j^{in}$ are unobserved individual fixed effects. To estimate coefficients, consider tetrads composed of four users, $i$, $j$, $k$, and $l$. The likelihood of $l$ follows $j$, conditional on either $l$ follows $j$ or $l$ follows $k$, could be shown that it does not depend on $\alpha_k^{out}$:
 \begin{equation}
 \nonumber
P(W_{lj}=1|D,X,\alpha,W_{lj}+W_{lk}=1)= \frac{exp\left[(D_{lj}-D_{lk})\beta_D+(X_{lj}-X_{lk})^T\beta_X+\alpha_j^{in}+\alpha_k^{in}\right]}{1+exp\left[(D_{lj}-D_{lk})\beta_D+(X_{lj}-X_{lk})^T\beta_X+\alpha_j^{in}+\alpha_k^{in}\right]}
\end{equation}
Similar derivations could be done for events $P(W_{lj}=1|W_{lj}+W_{lk}=1)$, $P(W_{lj}=1|W_{lj}+W_{ij}=1)$, etc., and the conditional likelihood of $l$ follows $j$ does not depend on any fixed effect:
\begin{equation}
\nonumber
P(W_{lj}=1|D,X,\alpha,W_{lj}+W_{lk}=1,W_{ij}+W_{ik}=1,W_{lj}+W_{ij}=1)=
\end{equation}
\begin{equation}
\nonumber
\frac{exp\left[((D_{lj}-D_{lk})-(D_{ij}-D_{ik}))\beta_D+((X_{lj}-X_{lk})-(X_{ij}-X_{ik}))^T\beta_X\right]}{1+exp\left[((D_{lj}-D_{lk})-(D_{ij}-D_{ik}))\beta_D+((X_{lj}-X_{lk})-(X_{ij}-X_{ik}))^T\beta_X\right]}
\end{equation}
Although few data points would be used for estimation, the model does not induce the incidental parameter problem and works well in sparse settings. We apply the method to both the directed network and the friendship network, then compare the coefficients with that before adding user fixed effects into the models.

\begin{figure} \centering
  \begin{minipage}[t]{0.45\textwidth}
    \centering
    \includegraphics[width=9cm]{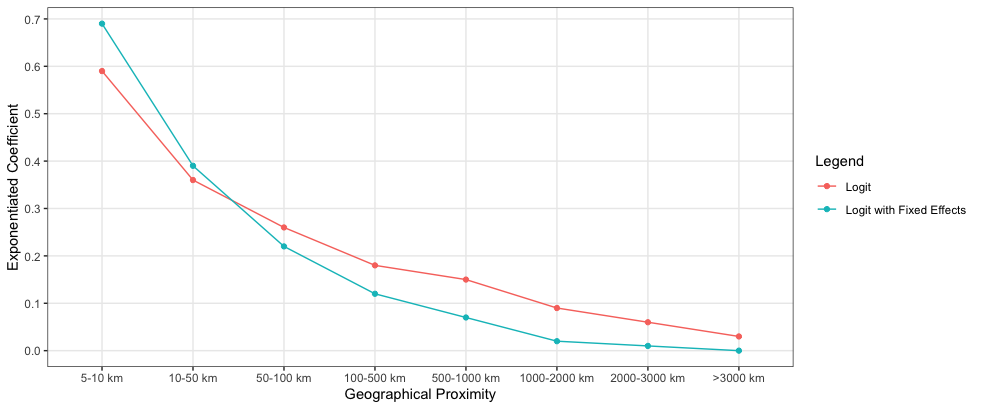}
  \end{minipage}
  \begin{minipage}[t]{0.45\textwidth} \centering
    \includegraphics[width=9cm]{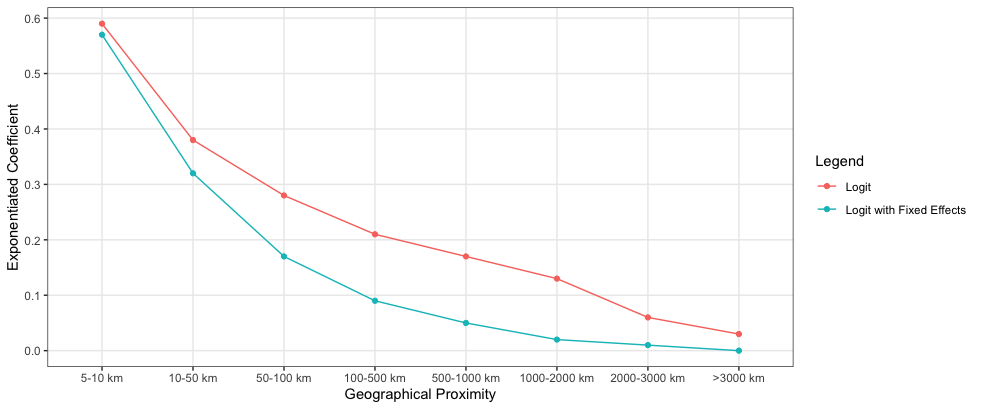}
  \end{minipage}
  \caption{Exponentiated Coefficients of Directed (left) and Friendship (right) Network with User Fixed Effects}
\end{figure}

For the friendship network, omitting unobserved characteristics makes the Logit model overstate the odds for all proximity levels. The error is larger for users who are between 50 and 2,000 kilometers, and smaller for users who live in the same place (within 50 kilometers) or beyond 2,000 kilometers. For directed network, the error is generally smaller, and the Logit model would underestimate the effects for users from the same place, but overestimate if the distance between two users is further than 50 kilometers.

\section{Conclusion}

In this paper, we study the role of geographical proximity with respect to link formation in a global online social network with dyadic logit models. We test on three hypotheses: people still prefer to connect with others nearby in a virtual world; users like to connect with those who come from popular countries; users from different countries may exhibit varying behaviors in sending friend requests. Data supports the first and the third hypothesis but has not found significant evidence for the second hypothesis.

Generally, the probabilities to form connections decrease with the rising physical distance between two users, and how they decay depends on users' countries of origin. Proximity is crucial for link formation when the two users are within meeting distance, but becomes weaker if they live in different cities or countries. It also has a lesser effect on a network composed of strong ties, or mutual connections, relative to a network with weak ties, or unidirectional connections.

It is important to note that the potential confounding factor of the app's friend recommendation algorithm, which could also be location-based, could contaminate the empirical results. Isolating the effect of geographical proximity is challenging using the current data, and a few strategies may be adopted when additional information is available. For example, additional control variables for the recommendation algorithm could be included provided the logic of the algorithm is known. Identifying instrumental variables that affect physical distance but are unrelated to the recommendation algorithm is also helpful, e.g. travel time between the two users' locations, policy changes that impact the layout or accessibility of certain areas, etc.

Apart from the potential confounding factors, there is still room left for future work. Firstly, from the graph topology, it can be seen that the network is composed of many regular-type users with fewer connections and a few celebrity users with a great number of followers. There may be reasons to believe that regular type users and celebrity type users behave differently in making connections and thus should be modeled separately. Thirdly, the topology of the network has not been fully taken advantage of in the current model. New variables that reflect both local and global network structure, such as graphon, could be added into the model to capture users’ structural tastes for the network, just to name an example. Fourthly, in the present study, all possible dyads are considered, and the assumption is that each user goes over all 11,992 profiles and selects whom to follow. However, users may not be aware of at least some groups of people, e.g. they may only evaluate others who show up in their feeds unless they intentionally search all users worldwide. Thus it is reasonable to develop a model that assumes the users make their follow-unfollow decisions based on some limited consideration sets instead of the entire population. Lastly, it would be interesting to see how social networks in different countries evolve each day, by analyzing more daily snapshots within an observation window and comparing the degree of homophily versus transitivity, as in \cite{graham2016homophily}.

\newpage
\bibliographystyle{unsrt}  
\bibliography{references}

\end{document}